\documentclass{aa}  

\usepackage{graphicx}
\graphicspath{{./}{figures/}}
\usepackage{xcolor}
\usepackage{txfonts}
\usepackage{hyperref}
\usepackage[caption=false]{subfig}

\begin{document} 

\nolinenumbers

\title{Observation of solar radio burst events from Mars orbit with the Shallow Radar instrument}

   \author{Christopher Gerekos
          \inst{1}\fnmsep\thanks{Corresponding author,\\\email{christopher.gerekos@austin.utexas.edu}}
          \and
          Gregor Steinbrügge\inst{2}
          \and
          Immanuel C. Jebaraj\inst{3}
          \and
          Andreas Casillas\inst{4}
          \and
          Elena Donini\inst{5}
          \and
          Beatriz S\'anchez-Cano\inst{6}
          \and
          Mark Lester\inst{6}
          \and
          Jasmina Magdalenić\inst{7}\fnmsep\inst{8}
          \and
          Sean T. Peters\inst{9}
          \and
          Andrew Romero-Wolf\inst{2}
          \and
          Donald D. Blankenship\inst{1}
          }

   \institute{University of Texas at Austin Institute for Geophysics, J.J. Pickle Research Campus, 10100 Burnet Road, 78758 Austin, TX, USA.
    \and
    Jet Propulsion Laboratory, California Institute of Technology, 4800 Oak Grove Drive, CA 91011, USA.
    \and
    Department of Physics and Astronomy, University of Turku, Turku, Finland.
    \and
    Naval Postgraduate School, Monterey, CA, United States.
    \and
    Fondazione Bruno Kessler, Via Sommarive 18, 38123 Povo, Trento, Italy
    \and
    School of Physics and Astronomy, University of Leicester, University Rd, Leicester LE1 7RH, UK
    \and
    Center for mathematical Plasma Astrophysics, Department of Mathematics, Katholieke Universiteit Leuven, Celestijnenlaan 200B, B-3001 Leuven, Belgium
    \and
    Solar-Terrestrial Centre of Excellence, Royal Observatory of Belgium, Avenue Circulaire 3, 1180 Uccle, Belgium
    \and
    Department of Aerospace Engineering Sciences, College of Engineering and Applied Sciences, University of Colorado Boulder, CO, USA
             }

   \date{Accepted 19/12/2023}

  \abstract
   {Multispacecraft and multiwavelength observations of solar eruptions such as flares and coronal mass ejections are essential to understand the complex processes behind these events. The study of solar burst events in the radio-frequency spectrum has relied almost exclusively on data from ground-based observations and a few dedicated heliophysics missions such as STEREO or Wind.}
   {Reanalysing existing data from the Mars Reconnaissance Orbiter (MRO) Shallow Radar (SHARAD) instrument, a Martian planetary radar sounder, we have discovered the instrument was also capable of detecting solar radio bursts, and was able to do so with unprecedented resolution for a space-based solar instrument. In this study we aim at demonstrating the reliability and value of SHARAD as a new solar radio-observatory.}
   {We characterised the sensitivity of the instrument to type-III solar radio bursts through a statistical analysis of correlated observations, using STEREO and Wind as references. Using 38 correlated detections, we establish the conditions under which SHARAD can observe solar bursts in terms of acquisition geometry. As an example of scientific application, we also present the first analysis of type-III characteristic times at high resolution beyond 1 AU.}
   {A simple logistic model based purely on geometrical acquisition parameters can predict burst show vs. no-show in SHARAD data with an accuracy of 79.2\%, demonstrating the reliability of the instrument for detecting solar bursts and laying the foundation for using SHARAD as a solar radio-observatory. The extremely high resolution of the instrument, both in temporal and frequency directions, its bandwidth, and its position in the solar system enable SHARAD to make significant contributions to heliophysics; it could inform on plasma processes on the site of the burst generation and along the propagation path of associated fast electron beams.}
   {}

   \maketitle

\section{Introduction}\label{sec1}

The Sun is routinely capable of accelerating electrons to suprathermal energies through eruptive phenomena such as flares, coronal mass ejections (CMEs), and the subsequent shock waves that they drive. Through various mechanisms, these suprathermal electrons emit radiation in the entire electromagnetic spectrum and particularly in the longer radio wavelengths. Based on the source (streaming electrons, shock waves, etc.) the emission can manifest itself with different morphological properties on the dynamic radio spectrogram. Early observations have distinguished five main spectral types \citep[I--V;][]{Kundu1965}, and further sub-classifications have been made since.

In the past two decades, radio frequency observatories onboard heliophysics missions such as the Solar TErrestrial RElations Observatory Ahead \& Behind \citep[STEREO A \& B/WAVES][]{Bougeret08}, and Wind/WAVES \citep{Bougeret95} have been used to understand various aspects of interplanetary radio emissions. Recently, they have been combined with the Radio Frequency Spectrometer \citep[RFS, part of the FIELDS suite;][]{Bale16, Pulupa17} onboard the Parker Solar Probe \citep[PSP;][]{Fox16} and the radio and plasma waves \citep[RPW;][]{Maksimovic20b} instrument onboard Solar Orbiter \citep[SolO;][]{Muller13} for multi-vantage point studies of hecto-kilometric (H-K) radio emissions \citep[e.g.][]{Jebaraj23b, Dresing23}. While they do not provide the same time-frequency resolution as ground-based instrumentation, observing radio emissions simultaneously from multiple vantage points presents the possibility to investigate the generation and propagation of radio waves \citep{Musset21}. Previous multi-vantage point observations of solar burst events made beyond 1 AU include that of \cite{lecacheux1989characteristics} and \cite{bonnin2008directivity}. 

Planetary radar sounders are a class of spacecraft-mounted remote sensing instruments that operate by recording reflections of electromagnetic waves off a solid planetary body. Such reflections arise when an incoming electromagnetic field encounters a change in the dielectric constant of the medium, such as the space-surface interface, subsurface layering, or subsurface inclusions. The source of this incoming field can be the instrument itself, in which case the radar will transmit a coded waveform (usually a linear chirp) with a power of a few Watts. This mode of operation is known as \emph{active sounding} \citep{skolnik1980introduction}. Conversely, the incoming field can be a signal of opportunity of astrophysical origin, a recently-proposed mode of operation known as \emph{passive sounding} \citep{romero2015passive}. Radar sounders typically operate in the deca-hectometre (D-H) wavelength bands.

A particularly productive planetary radar sounder has been the Shallow Radar (SHARAD) instrument \citep{seu2004sharad} onboard the National Aeronautics and Space Administration (NASA) Mars Reconnaissance Orbiter (MRO) mission, which was launched towards Mars in 2005 and began observations a year later. SHARAD is sensitive in the 13.3 to 26.7 MHz band, has a time resolution of 1.43 ms before pre-summing, and a frequency resolution of 7.41 kHz \citep{seu2004sharad, croci2011shallow}. The radiating element of SHARAD is a thin-wire 10 m long dipole antenna \citep{croci2011shallow}. When operating SHARAD, the spacecraft is nominally oriented so that its -Z direction points towards the Martian surface, although some variations in the roll angle have been considered to accommodate unforeseen lobe distribution of the SHARAD antenna pattern \citep{croci2011shallow,campbell2021calibration}. Amongst other discoveries enabled by SHARAD, \cite{grima2009north} characterised the ice purity in the layers of the Martian polar caps, \cite{holt2008radar} found evidence of buried glaciers at mid-latitudes, and \cite{campbell2013roughness} constrained the roughness and near-surface density of the Martian surface. 

In this manuscript we show how SHARAD can be used to observe solar radio bursts events at unprecedented resolution in the D-H band from space, also marking the first attempt to use a planetary radar sounder as a solar radio-observatory. We note that \cite{gurnett2010non} mention having seen solar radio bursts in data from the Mars Advanced Radar for Subsurface and Ionosphere Sounding (MARSIS), another Martian radar sounder \citep{marsis} operating at markedly lower frequencies than SHARAD (1.3 to 5.5 MHz in sounding mode), but these were treated as parasitic signals, and no attempt has yet been made to study solar radio burst with this instrument. In order to demonstrate the potential of SHARAD for heliophysics, we devised a purpose-built algorithm for detection of SHARAD-detectable type III bursts in the STEREO/WAVES and Wind/WAVES datasets. We propagated these events in time from the orbits of these spacecraft to the orbit of Mars, and looked up for any instance when the SHARAD receiver was on and when the Sun was not occulted by Mars as seen from MRO. This search for SHARAD type III burst-containing candidates yielded 179 distinct SHARAD radargrams, of which 38 contain solar burst. Through a multivariate statistical analysis of the candidates, we quantify the importance of the acquisition geometry for burst detection. As, by construction, all our SHARAD-detected bursts have at least one correlated observation, we also present comparisons of a few selected bursts observed by SHARAD and the source solar spacecraft, and provide a succinct commentary on their features. As an example of scientific application of this dataset, we analyse the characteristic times of our dataset of type-III bursts following a methodology close to that of \cite{reid2018solar} and compare the frequency-dependence scaling laws we obtain with that obtained on Earth.

This manuscript is structured as follows. Section \ref{sec:methods} presents our methodology for finding correlated observation opportunities and building the proposed list of SHARAD candidates. Section \ref{sec:obs} summarises our observations. Section \ref{sec:sensitivity} presents the statistical analysis of the sensitivity of SHARAD to type III SRBs. Section \ref{sec:disc} contains in-depth analyses of representative SHARAD spectrograms, including the type III characteristic time analysis at 1.5 AU. Section \ref{sec:concl} concludes this paper with a discussion of the proposed dataset and of its perspectives. Appendix \ref{sec:dutycyle} presents a SHARAD duty cycle analysis and Appendix \ref{sec:quicklook} contains a quicklook mosaic of all the SHARAD-captured bursts we have found.

\section{Methods and datasets} \label{sec:methods}
In this section we present the methodology we used to detect solar radio bursts in STEREO and Wind data and to search for corresponding SHARAD candidates, focusing on type III bursts. The critical parts of the code have been made available along with an accompanying flowchart as Supplementary Material 1 \citep{supplmat1}.

\subsection{Solar radio bursts detection from STEREO and Wind}\label{sec:solarmethods}
Data from the STEREO and Wind iterations of the WAVES instrument was used for the compilation of SHARAD-detectable type-III bursts for a number of reasons:  the availability of round-the-clock observations, the existence of some degree of homogeneity of the data properties across STEREO and Wind, and the partial bandwidth overlap with SHARAD. These properties ensure that a relatively unified algorithm can be built to search for bursts in three different spacecraft and collect a large number of co-observation opportunities.

We used the the 60 seconds-averaged WAVES products from both STEREO and Wind, covering a period starting on 6 December 2006 (first SHARAD acquisition) and ending on 31 December 2021. An initial screening for bursts is made by integrating the spectrograms over their entire range of frequencies, yielding a one-dimensional time-series of power flux, and by detecting prominences that are one standard deviations higher than the background through the MATLAB proprietary function \texttt{findpeaks}. This is done for each 24h-spanning data product. A 3-step fuzzy logic algorithm is then applied on these peaks in order to reject prominences due to interferences or instrument malfunction and to reject genuine type III bursts that do not have significant energy above 13 MHz, about the lowest frequency available to SHARAD. To this end, the algorithm tests peaks flatness, peak asymmetry, and high-frequency content. For more details please refer to \citep{supplmat1}.

Given the large number of type III bursts that these missions recorded, and given the nature of our study, the algorithm is rather restrictive as false positives are more dangerous than false negatives. With this method we detected 5676 bursts with STEREO A, 4340 bursts with STEREO B, and 3538 events with Wind.

\begin{figure}[t]
    \centering
    \includegraphics[width=0.45\textwidth]{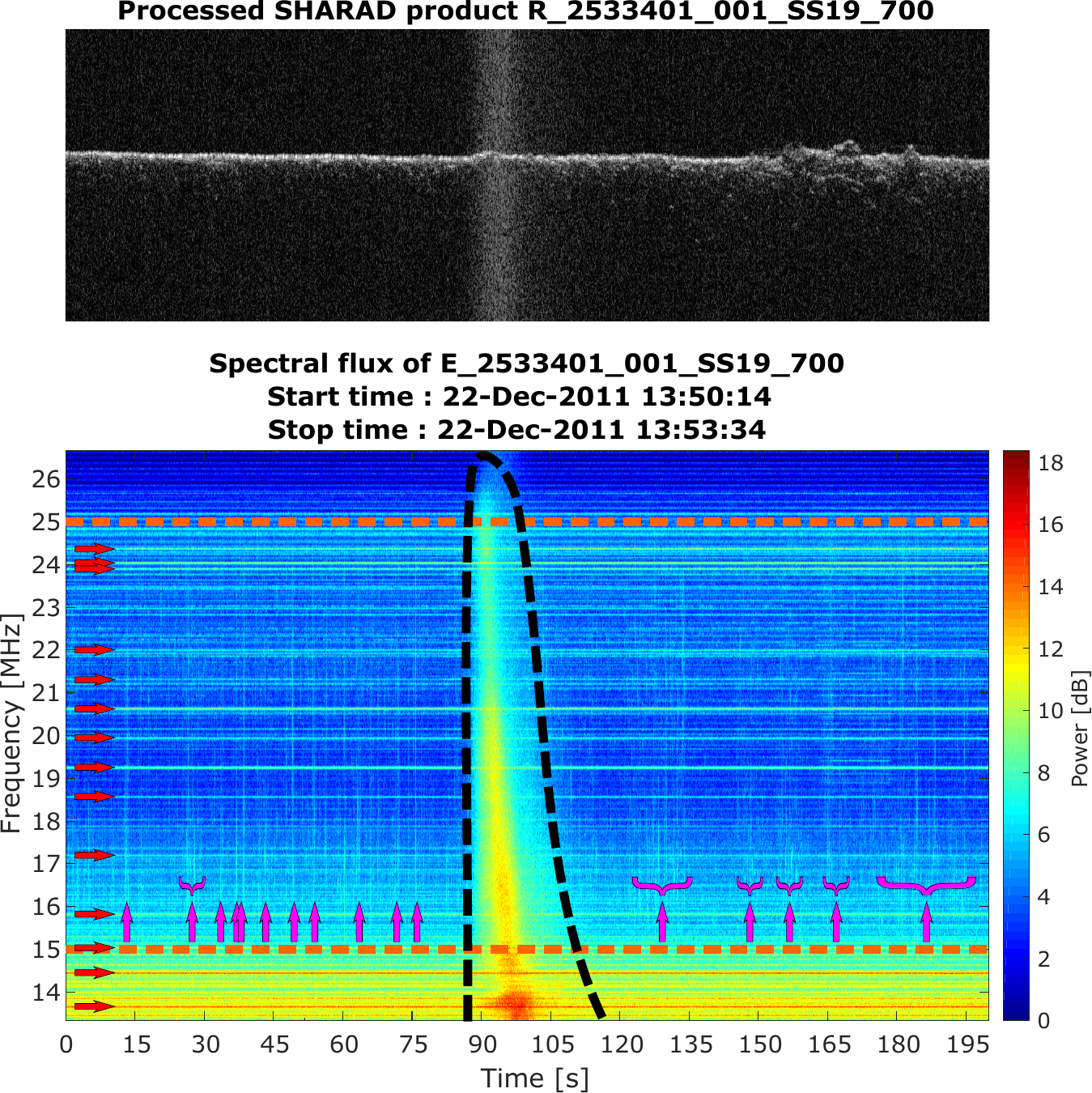} 
    \caption{Comparison of the Reduced Data Record (top) and spectrogram of the Experimental Data Record (bottom) for SHARAD product 2533401, along with annotations for the main signals present in the spectrogram. This product contains a type III burst outlined by the dashed black curve ; it is affected by various EMI sources, which are the narrow band signals that are persistent in time and manifest themselves as horizontal traces, the strongest of which are highlighted with horizontal arrows ; and it contains traces of the reflection of the active chirp, which are are wideband signals with short durations in time and manifest themselves as vertical traces, the strongest of which (or groups of which) are highlighted by vertical arrows. The band of flat spectral gain is delimited by the orange dashed lines at 15 and 25 MHz.}
    \label{fig:workedex}
\end{figure}

\subsection{SHARAD candidate selection}  \label{sec:sharad-methods}
Each of those events has an associated timestamp $t$ that corresponds to the time of detection of peak power. We propagate this time of detection at the observer (STEREO A, B, or Wind) to Martian orbit by considering the radial distance separating the solar spacecraft and Mars at that time. This rests on the hypothesis that the isocontours of burst detection time form circles around the Sun. This radial distance difference is converted into a delay $\tau$ assuming propagation at the speed of light in vacuum. In other words, if $t'$ is the supposed time of detection at Mars, we can write
\begin{equation}
    t' = t + \tau = t+ c^{-1} \left(\left| \mathbf{v}_M - \mathbf{v}_O \right| - \left| \mathbf{v}_{w} - \mathbf{v}_O \right|\right) = t+\frac{r_M - r_w}{c},
\end{equation}
where $\mathbf{v}_w$, $\mathbf{v}_O$ and $\mathbf{v}_M$ represent the position of the observer (STEREO A, B, or Wind), the position of the Sun, and the position of Mars, respectively, and where $r$ represents a radial distance from the Sun. These positions can be extracted from the appropriate SPICE kernels. The corresponding Mars detection time $t'$ is computed for all of the events detected in the previous step. Then, a SHARAD data product is considered a candidate if it satisfies the two following conditions: (i) SHARAD is operating at $t'$; and (ii) the Sun is not occulted by Mars at $t'$ as seen from MRO. The first criterion is verified by testing that the start and stop time $t_1$ and $t_2$ of a given SHARAD data product is such that $t_1 \le t' \le t_2$, where $t_1$ and $t_2$ are found in the SHARAD product metadata. The second criterion can be easily tested using a SPICE query.

After application of these criteria, we are left with 226 SHARAD detections comprising 179 distinct SHARAD data products (a given event recorded by different solar observatories may correspond to the same SHARAD product). This very strong reduction of the numbers of events, compared with that of Section \ref{sec:solarmethods} is explained by the very low duty cycle of SHARAD activity (see Appendix \ref{sec:dutycyle}) and by the fact a majority of radar sounding observations were done on the Martian nightside \citep{campbell2013sharad}. A complete list of these SHARAD candidates is given as Supplementary Material 2 \citep{supplmat2}.

\subsection{Spectrogram and time-profile generation}
The SHARAD Experimental Data Records \citep[EDR;][]{edrsis} are real-valued baseband waveforms containing 3600 samples per rangeline. A rangeline is a 1D fast-time collection of samples that forms the basic acquisition unit of a radargram, and a radargram is a collection of consecutive rangelines. On SHARAD, rangelines are acquired at a frequency of 700 Hz, and have a duration of 135 $\mu$s each. The SHARAD dynamic spectra shown in this manuscript are the absolute value of the fast Fourier transform of the EDR product in dB scale. Only the positive-frequency components are displayed (1800 samples). In addition to possible solar radio bursts, these opportunistically-acquired SHARAD SRB spectrograms contain a number of signals of non-solar origin: electromagnetic interference (EMI) and reflections of the active chirp. These are highlighted in Figure \ref{fig:workedex} for product 2533401, which contains a type III burst. For canonical exploitation of radar data, that is, the geologic analysis of backscattered echoes, the EDR data must at least undergo an operation known as \emph{range-compression} to be interpretable \citep{cumming2005digital, gerekosth}. On SHARAD, range-compressed Experimental Data Records are known as Reduced Data Records \citep[RDR;][]{rdrsis} and constitute the "usual" form of SHARAD data. The corresponding RDR product for 2533401 is also shown in Figure \ref{fig:workedex} for context. It depicts a slightly rough surface located in the martian northern lowlands, with a small crater near the end of the acquisition. The solar burst is clearly visible as the noisy part in the centre.

It must be noted that SHARAD has not been calibrated on the ground, meaning there is no formally-generated data products in physical units of spectral flux \citep{rdrsis}. The gain per frequency band across the whole bandwidth (13.3 to 26.7 MHz) has also not been completely characterised, although it was optimised to be spectrally-flat in the 15 to 25 MHz range \citep{shafum}. Calibration of SHARAD \emph{a posteriori} is an ongoing effort \citep{campbell2021calibration, castaldo2013scientific}. 

No gain de-trending or active chirp removal algorithms have been applied to the spectrograms shown in this work, as our exploitation of the data in this letter remains exploratory. Instead, a basic filtering using a Gaussian kernel of 3 pixels is applied globally for image rendition concerns. The time-profiles represent a 100-pixel moving average of the raw spectra.

\section{Observations} \label{sec:obs}
The 179 SHARAD candidates obtained from our search algorithm were visually examined for the presence of solar bursts. Of these candidates, 38 products contain a clear, scientifically-exploitable burst. The list of all candidates, including those without a burst, has been made available as Supplementary Material 2 \citep{supplmat2} to allow for possible future reanalyses that would look for more subtle signatures, and the subset of SHARAD candidates which contain an unambiguous solar burst is summarised in Table \ref{tab:yesbursts}. A quicklook mosaic of these products is further provided in Appendix \ref{sec:quicklook}.

\begin{figure*}[h!t]
    \centering
        \subfloat[]{\includegraphics[scale=0.25]{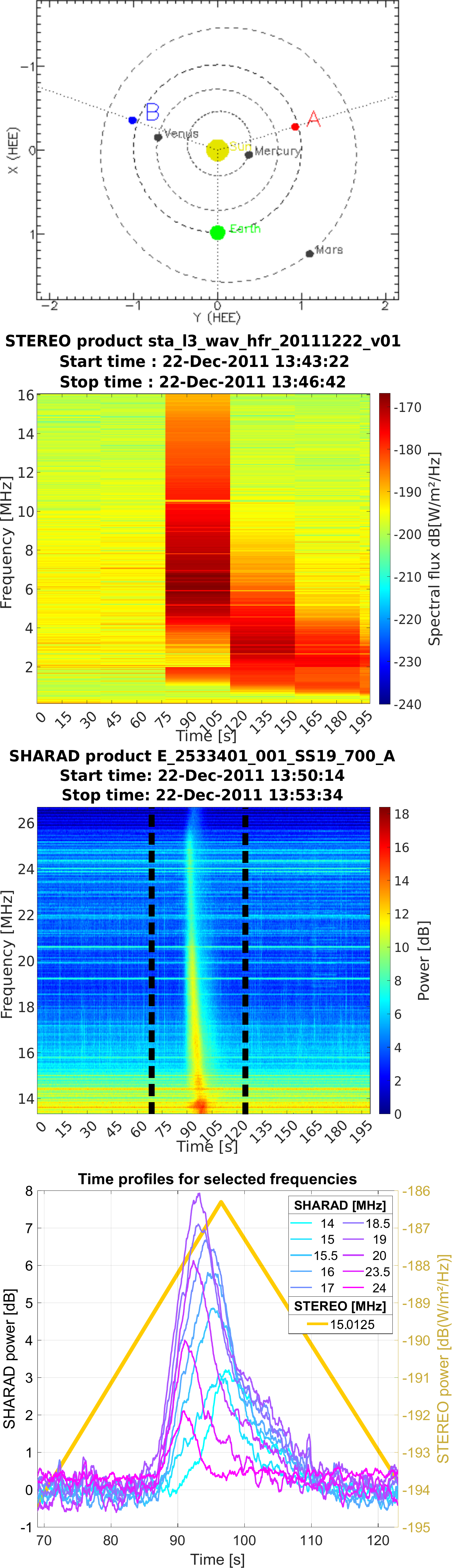}}\hfill
        \subfloat[]{\includegraphics[scale=0.25]{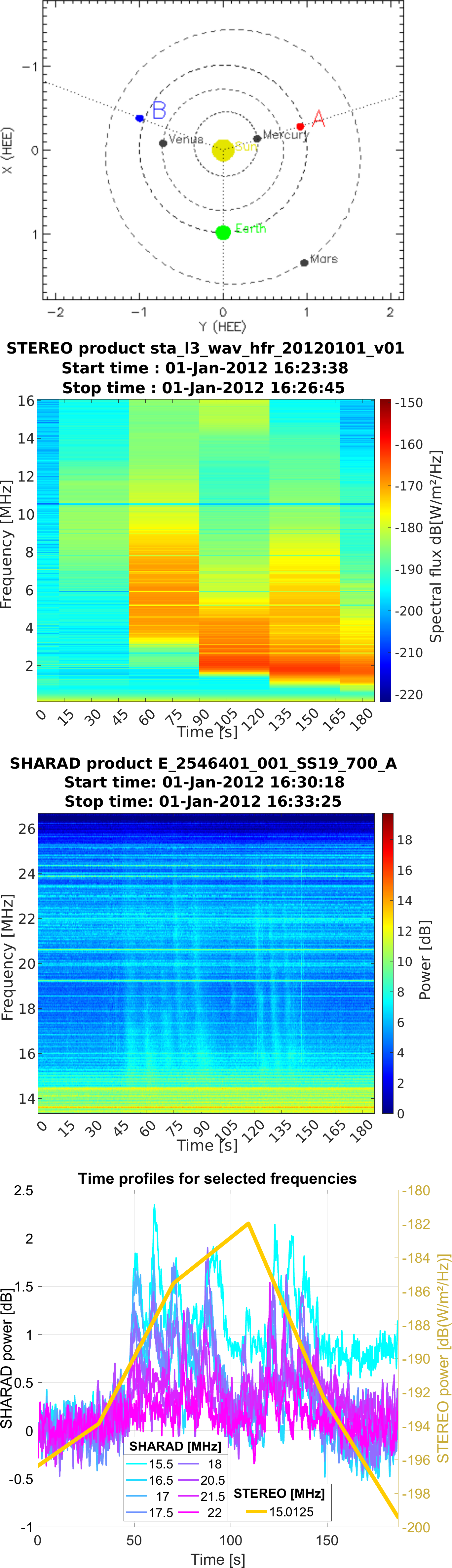}}\hfill
        \subfloat[]{\includegraphics[scale=0.25]{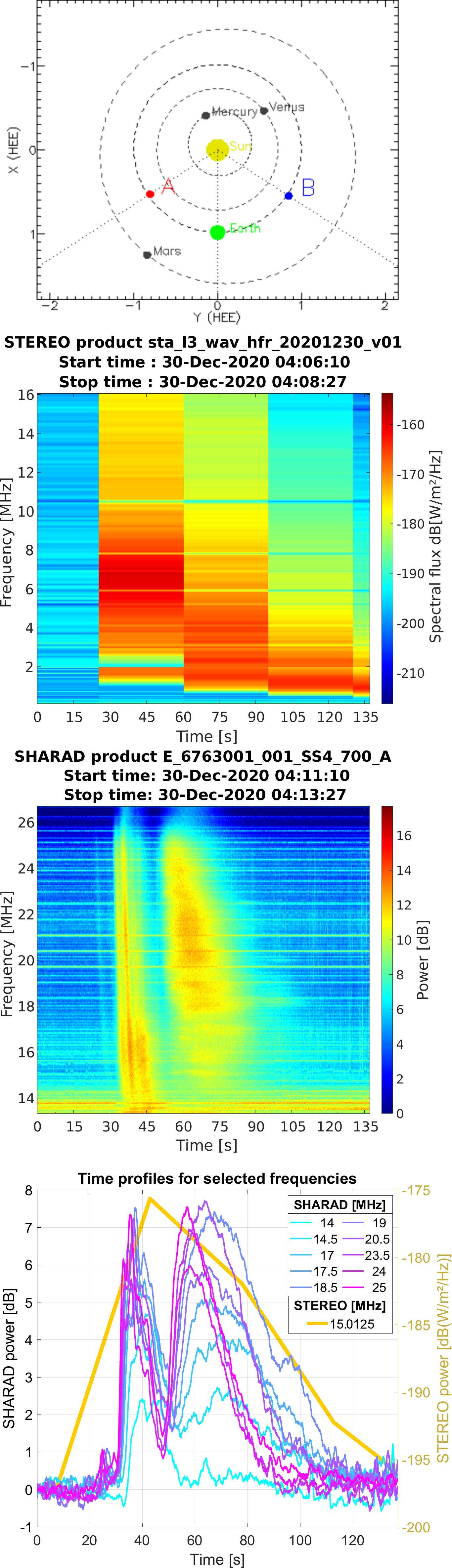}}
    \caption{In-depth analysis of three representative elements of the SHARAD burst dataset: (a) SHARAD product 2533401, which is also displayed in Figure \ref{fig:workedex}, (b) SHARAD product 2546401, (c) SHARAD product 6763001. For each case, the top plot represents a map of the solar system. The second row shows the STEREO-captured original burst, in calibrated units. The third row shows the SHARAD spectrogram. The bottom plot contain time-profiles for selected frequencies extracted from the SHARAD spectrogram, along with the 15.0125 MHz line from STEREO for comparison. In the case of 2533401, the dashed black lines on the SHARAD spectrogram show the bounds of the corresponding time-profiles. Calibrated STEREO data sourced from \citep{stereocal}.}
    \label{fig:examples}
\end{figure*}

Several of these bursts display extensive fine structure (see also Figure \ref{fig:examples}c), and some bursts have a peculiar dynamic spectra unlike the typical type III profile. Interestingly, a type-II event (SHARAD product 3544101) has also been picked up by our algorithm.

\begin{table*}
    \centering
    \begin{tabular}{c|c c c c c}
        Observation Source & SHARAD prod. & Date & ROI start (UT) & ROI stop (UT) & Notes\\
        \hline \hline
        STEREO B            & 1301501 & 06 May 2009 & 15:46:41 & 15:47:49 & pecul.\\
        STEREO A, B, WIND   & 2275901 & 04 Jun 2011 & 22:03:17 & 22:05:11 & pecul.\\
        STEREO A            & 2387602 & 30 Aug 2011 & 22:35:50 & 22:41:04 & * \\
        STEREO A            & 2404001 & 12 Sep 2011 & 17:29:13 & 17:30:05 & \\
        STEREO A            & 2433801 & 05 Oct 2011 & 23:00:51 & 23:03:31 & \\
        WIND                & 2532101 & 21 Dec 2011 & 13:23:48 & 13:26:02 & \\
        STEREO A, WIND      & 2533401 & 22 Dec 2011 & 13:50:14 & 13:53:34 & * \\
        WIND                & 2545801 & 01 Jan 2012 & 05:28:47 & 05:31:50 & * \\
        STEREO A            & 2546401 & 01 Jan 2012 & 16:30:18 & 16:33:25 & * \\
        STEREO B	        & 2712401 & 10 May 2012 & 00:14:07 & 00:17:09 & f.s.\\
        STEREO A, B, WIND   & 2714201 & 11 May 2012 & 09:56:01 & 09:58:18 & \\
        WIND                & 2746002 & 05 Jun 2012 & 05:36:38 & 05:39:30 & \\
        WIND                & 2760001 & 16 Jun 2012 & 03:20:14 & 03:22:31 & * \\
        STEREO B            & 2792501 & 11 Jul 2012 & 09:59:39 & 10:02:19 & \\
        WIND                & 2813001 & 27 Jul 2012 & 10:10:11 & 10:13:31 & * \\
        STEREO A, WIND      & 3397801 & 26 Oct 2013 & 02:49:17 & 02:51:57 & \\
        STEREO A            & 3401901 & 29 Oct 2013 & 07:40:56 & 07:44:16 & \\
        STEREO A            & 3403202 & 30 Oct 2013 & 07:54:43 & 07:56:14 & \\
        STEREO A, WIND      & 3404001 & 30 Oct 2013 & 22:58:27 & 22:59:47 & * \\
        STEREO A, B         & 3487401 & 03 Jan 2014 & 22:21:02 & 22:23:31 & * \\
        STEREO A, B         & 3537001 & 11 Feb 2014 & 13:43:34 & 13:45:25 & pecul.\\
        WIND                & 3543602 & 16 Feb 2014 & 17:41:47 & 17:43:12 & \\
        WIND                & 3543702 & 16 Feb 2014 & 19:25:59 & 19:34:11 & * \\
        WIND                & 3544101 & 17 Feb 2014 & 02:55:44 & 03:02:24 & type-II\\
        STEREO A            & 3547101 & 19 Feb 2014 & 10:49:32 & 10:52:23 & * \\
        STEREO A, B         & 3564202 & 04 Mar 2014 & 18:36:40 & 18:48:23 & \\
        WIND                & 3582402 & 18 Mar 2014 & 22:51:36 & 22:55:41 & \\
        WIND                & 3617602 & 15 Apr 2014 & 09:21:50 & 09:22:35 & f.s.\\
        WIND                & 3649201 & 09 May 2014 & 23:52:30 & 23:55:22 & * \\
        STEREO A, B         & 3708701 & 25 Jun 2014 & 08:42:40 & 08:44:11 & f.s.\\
        STEREO B            & 3738401 & 18 Jul 2014 & 12:37:42 & 12:39:36 & \\
        STEREO A, B         & 3757101 & 02 Aug 2014 & 01:22:51 & 01:26:06 & \\
        STEREO A, B         & 3766901 & 09 Aug 2014 & 16:51:31 & 16:56:17 & * \\
        WIND                & 3799601 & 04 Sep 2014 & 05:03:43 & 05:04:52 & * \\
        WIND                & 3929701 & 14 Dec 2014 & 12:58:33 & 13:00:39 & \\
        STEREO A            & 6709701 & 18 Nov 2020 & 15:40:02 & 15:42:08 & \\
        STEREO A, WIND      & 6712501 & 20 Nov 2020 & 19:37:31 & 19:38:51 & \\
        STEREO A, WIND      & 6763001 & 30 Dec 2020 & 04:11:10 & 04:13:27 & f.s.\\
    \end{tabular}
    \caption{Table of all the SHARAD candidates where a burst (type III unless otherwise indicated) was identified within the product. The ROI start and stop time refer to the region of interest of the SHARAD spectrograms, and correspond to the temporal bounds of the mosaic shown in Appendix \ref{sec:quicklook}. In the "Notes" column, "pecul." indicates a possibly different type of burst ; "f.s." signifies a type III burst with extensive frequency-domain fine structure; whereas the asterisk (*) denotes a product that was used in the time-profile analysis of \ref{sec:risefall}. See Supplementary Material 2 \citep{supplmat2} for the candidates where we did not notice a burst.}
    \label{tab:yesbursts}
\end{table*}

\begin{figure*}
    \centering
    \subfloat[]{\includegraphics[height=6.3cm]{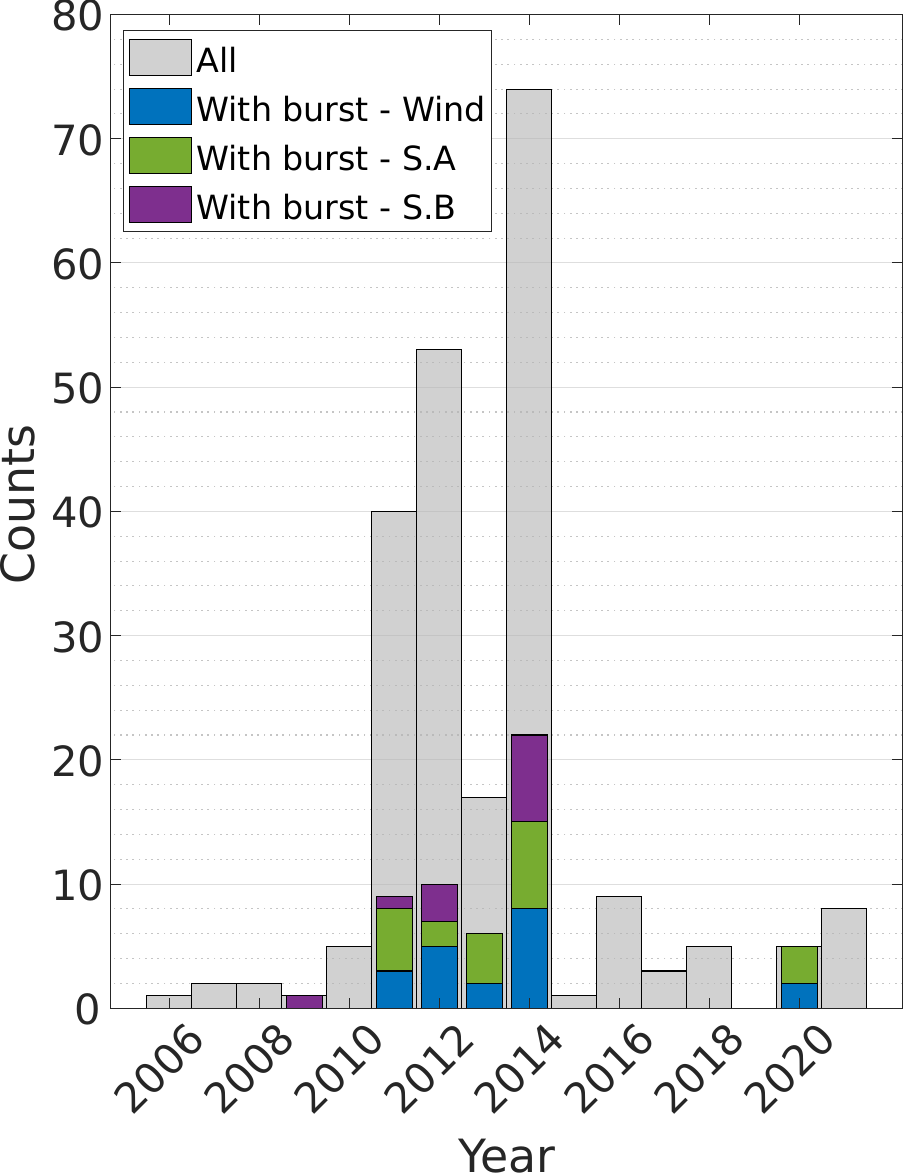}}\hfill
    \subfloat[]{\includegraphics[height=6.3cm]{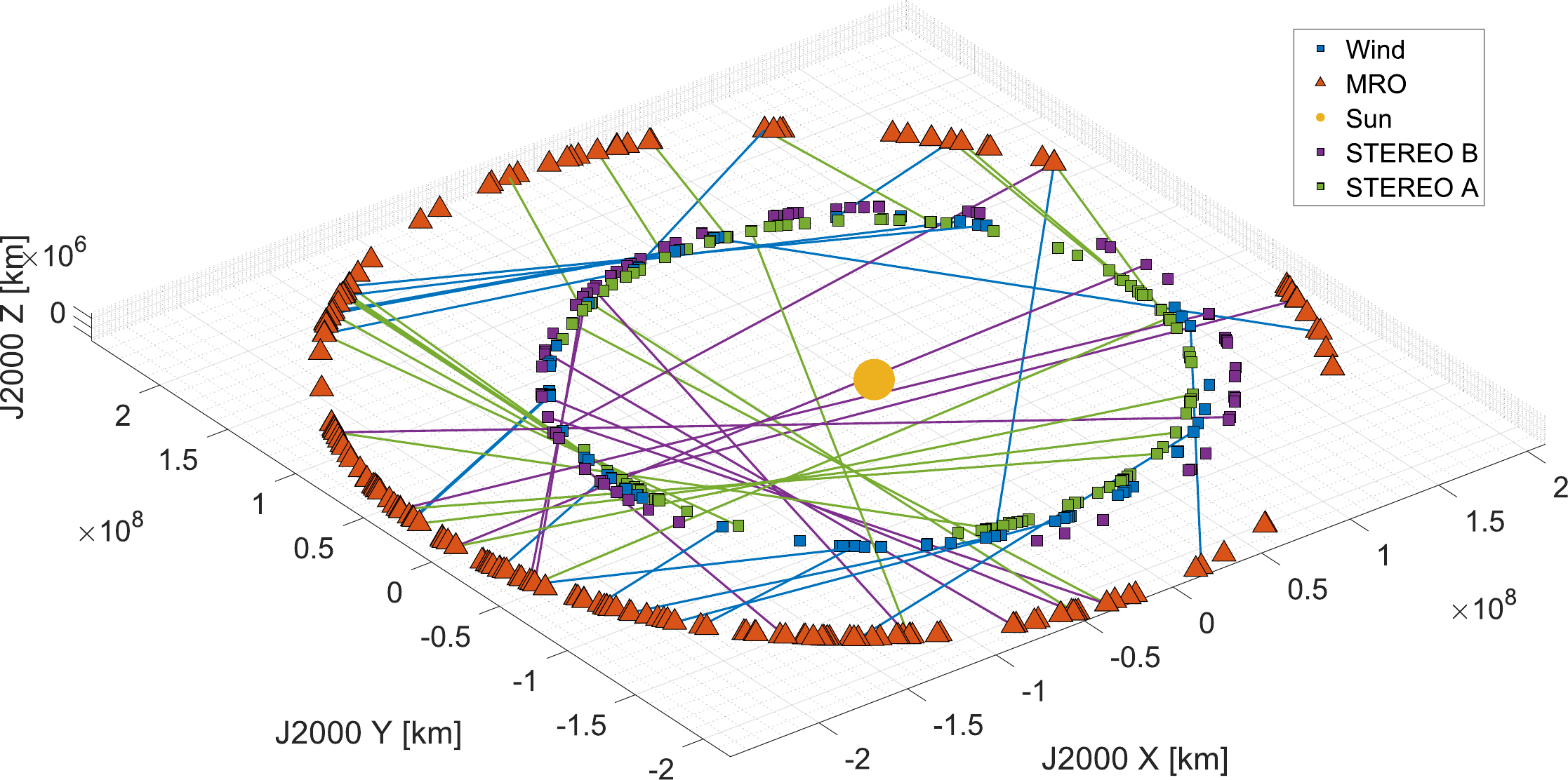}}
    \caption{Visual summary of the generated dataset. (a) Temporal distribution of the SHARAD burst candidates: count of all SHARAD candidates per year in gray, with the burst-containing subsets colour-coded according to the source of the observation. The histogram follows the solar cycle and the proportion of burst-containing products is stable. (b) Spatial distributions of SHARAD burst candidates: positions of MRO, STEREO A, STEREO B, and Wind for all SHARAD candidates. Lines are drawn between the source of the observation and the corresponding SHARAD candidate for every candidate with a burst.
    }
    \label{fig:generals}
\end{figure*}

The temporal and spatial distributions of the 226 SHARAD candidate events, with and without burst, is shown in Figure \ref{fig:generals}. The number of SHARAD candidates throughout the surveyed years closely follows the solar cycle, and the fraction of these candidates that were confirmed to contain a burst stays at around 25\% for every year. The additional variability in the number of candidates can be explained by the variation in the yearly SHARAD duty cycle (see Appendix \ref{sec:dutycyle}). This consistency suggests that the geometric selection criteria described in Section \ref{sec:sharad-methods} do not introduce any particular bias. When considering the spatial spread of the observations, it can be observed that the sources of the observations (STEREO A, B, and Wind) are rather evenly distributed whereas the corresponding SHARAD candidates are more concentrated on the negative X-side of the solar system in the J2000 system of coordinates. However, when considering the SHARAD candidates containing a burst, then the distribution of the sources shows a similar unbalance. The first feature is simply explained by the fact a great proportion of bursts were recorded in 2014 and that Mars was in that particular sector during that year, whereas the second observation hints at the angular difference dependence for detectability (see Section \ref{sec:sensitivity}): the dedicated solar missions may have been at any place in the solar system during the peak years, but it is those angularly closest to Mars that led to better chances of a SHARAD observation.

\section{SHARAD sensitivity to type III bursts} \label{sec:sensitivity}

Of the 179 SHARAD candidates picked up by the algorithm described in Section \ref{sec:methods}, 38 contain a prominent burst, a fraction of 21.23\%, and 141 were identified as not containing a burst. In this section we shall discuss these figures and quantify the effects of several geometrical factors on this cross-detection rate. 

\begin{figure}[t]
    \centering
        \includegraphics[width=0.4\textwidth]{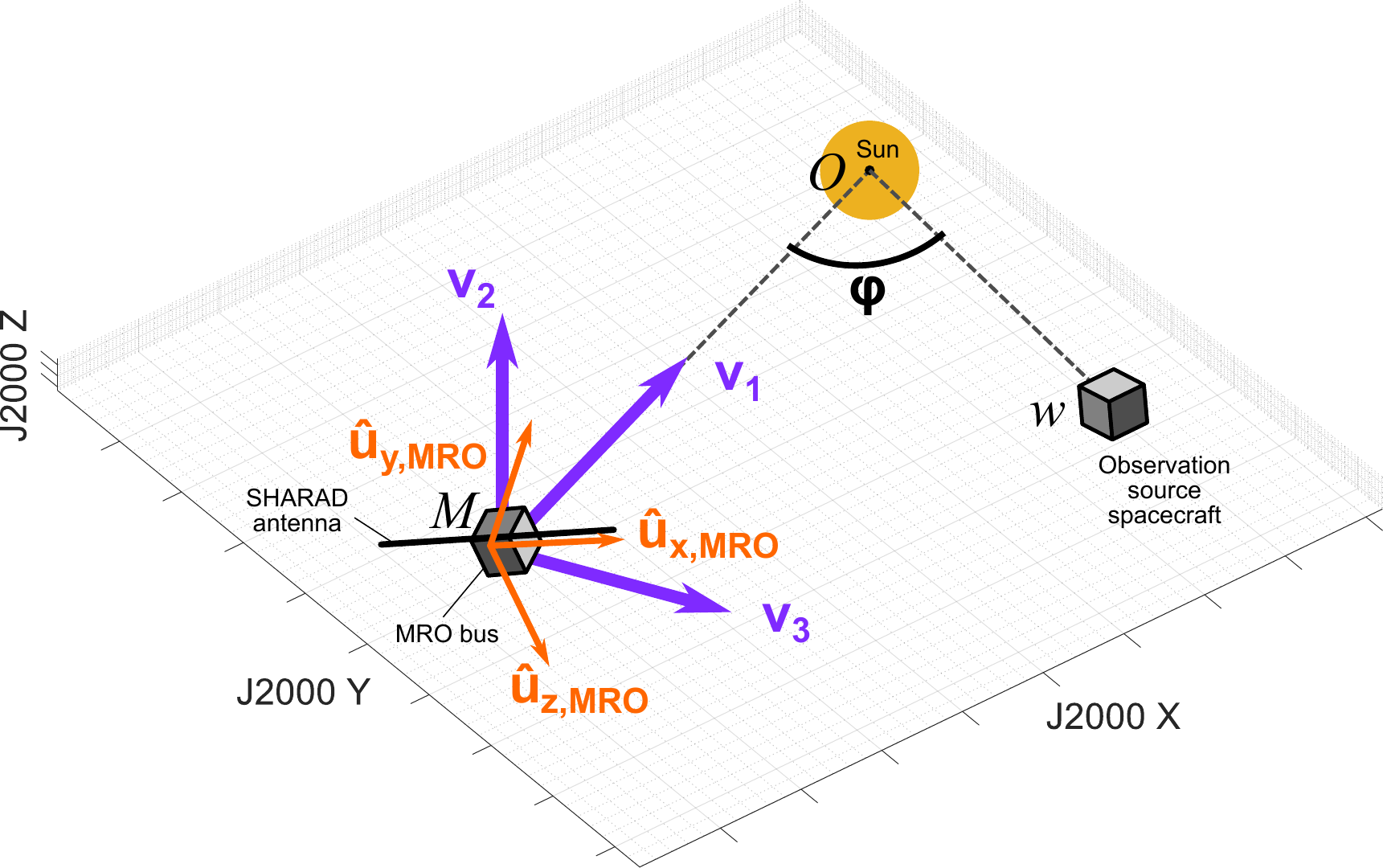}
    \caption{Illustration of several key quantities in the definition of relevant angles. In this illustration, yaw is about $60^\circ$, pitch is about $20^\circ$, and roll is close to $90^\circ$}
    \label{fig:anglesschem}
\end{figure}

An parameter that likely influences the detection of solar radio bursts with SHARAD is the absolute power of the burst as detected by the source of observations (STEREO A, B, or Wind): some bursts seen by the observer may not have been seen by SHARAD because of its own sensitivity threshold. However, it is not a parameter we can uniformly analyse, Wind/WAVES data has not yet been formally calibrated at the time of the analysis and issues of cross-calibration of STEREO/WAVES and Wind/WAVES are still unsettled (Krupar, pers.\ comm., 2022). Parameters of geometrical nature, however, can be analysed uniformly. The non-isotropic nature of type III bursts \citep[][]{Musset21} is expected to lead to a preferential detection if the angle between the source spacecraft and MRO with respect to the Sun (marked as $\phi$ in Figure \ref{fig:anglesschem}) is small. It is also expected that the orientation of MRO plays a major role in the detectability of bursts due to the antenna pattern of SHARAD \citep{croci2007calibration}. In processed radar sounding data, variations of gain of up to 4 dB have been observed by \cite{campbell2021calibration}. The position of the high-gain antenna (HGA) and the orientation of the solar panels of MRO are also known to considerably affect the antenna pattern of SHARAD \citep{campbell2021calibration}. Due to the complexity and interdependence of these effects, however, we chose to focus on the orientation of the antenna only. 

Since the orientation towards the Sun is the most relevant parameter, we shall define pitch, yaw, and roll angles with respect to that direction as follows. We first define a set of three reference unit vectors:
\begin{eqnarray}
        \mathbf{\hat{v}}_1 &=& \frac{\mathbf{v}_O - \mathbf{v}_{MRO}}{\left|\mathbf{v}_O - \mathbf{v}_{MRO}\right|},\\
        \mathbf{\hat{v}}_2 &=& \mathbf{\hat{v}}_\mathrm{z,ECLIPJ2000},\\
        \mathbf{\hat{v}}_3 &=& \frac{\mathbf{\hat{v}}_1 \times \mathbf{\hat{v}}_2}{\left| \mathbf{\hat{v}}_1 \times \mathbf{\hat{v}}_2 \right|},
\end{eqnarray}
where $\mathbf{v}_O = (0,0,0)$ is the position of the Sun in the J2000 system of coordinates, and $\mathbf{v}_{MRO}$ that of MRO. By construction, $\mathbf{\hat{v}}_1$ always points from MRO towards the Sun. $\mathbf{\hat{v}}_2$ is the normal to the ecliptic plane, and $\mathbf{\hat{v}}_3$ completes the orthonormal basis. With these vectors, we define the following projections of the $\mathbf{\hat{u}}_{j,MRO}$, $(j=x,y,z)$ vectors, which form the basis of MRO-fixed orthonormal unit vectors \citep{croci2007calibration}:
\begin{eqnarray}
        \mathbf{w}_x &=& \mathbf{\hat{u}}_{x,MRO} - (\mathbf{\hat{u}}_{x,MRO}\cdot\mathbf{\hat{v}}_3)\mathbf{\hat{v}}_3,\\
	\mathbf{w}_x' &=& \mathbf{\hat{u}}_{x,MRO} - (\mathbf{\hat{u}}_{x,MRO}\cdot\mathbf{\hat{v}}_2)\mathbf{\hat{v}}_2,\\
	\mathbf{w}_z &=&  \mathbf{\hat{u}}_{z,MRO} - (\mathbf{\hat{u}}_{z,MRO}\cdot\mathbf{\hat{v}}_3)\mathbf{\hat{v}}_3.
\end{eqnarray}
In Figure \ref{fig:anglesschem}, the vectors $\mathbf{\hat{v}}_i$, ($i=1,2,3)$ and $\mathbf{\hat{u}}_{j,MRO}$, $(j=x,y,z)$ are represented in purple and orange, respectively. $\mathbf{w}_x$ is the projection of $\mathbf{\hat{u}}_{x,MRO}$ onto the $(M,\mathbf{\hat{v}}_1, \mathbf{\hat{v}}_2)$ plane, $\mathbf{w}_x'$ is the projection of $\mathbf{\hat{u}}_{x,MRO}$ onto the $(M,\mathbf{\hat{v}}_1, \mathbf{\hat{v}}_3)$ plane, and $\mathbf{w}_z$ is the projection of $\mathbf{\hat{u}}_{z,MRO}$ onto the $(M,\mathbf{\hat{v}}_1, \mathbf{\hat{v}}_2)$ plane.

These constructions allow us to define the desired Sun-based pitch, yaw, and roll angles as follows:
\begin{eqnarray}
        \text{Yaw} &=& \text{sgn}(\mathbf{\hat{w}}_x'\cdot \mathbf{\hat{v}}_3)\arccos(\mathbf{\hat{w}}_x' \cdot \mathbf{\hat{v}}_1)\\
        \text{Pitch} &=&\text{sgn}(\mathbf{\hat{w}}_x\cdot \mathbf{\hat{v}}_2)\arccos(\mathbf{\hat{w}}_x \cdot \mathbf{\hat{v}}_1)\\
        \text{Roll} &=& \text{sgn}(\mathbf{\hat{w}}_z\cdot \mathbf{\hat{v}}_2)\arccos(\mathbf{\hat{w}}_z \cdot \mathbf{\hat{v}}_1)
\end{eqnarray}
where $\text{sgn}(\cdot)$ is the sign function. A $0^\circ$ value for both yaw and pitch depitcs a situation where $\mathbf{\hat{u}}_{x,MRO}$ is aligned with $ \mathbf{\hat{v}}_1$ and thus has the antenna axis pointed towards the Sun. An ideal dipole in this configuration would see zero gain in the direction of the Sun. A $90^\circ$ yaw angle places the antenna axis perpendicularly to the Sun in the horizontal plane (the projection $\mathbf{\hat{u}}_{x,MRO}$ unto the $(M,\mathbf{\hat{v}}_1, \mathbf{\hat{v}}_3)$ plane is perpendicular to $\mathbf{\hat{v}}_1$), whereas a $90^\circ$ pitch angle places the antenna perpendicularly to the Sun in the vertical plane (the projection $\mathbf{\hat{u}}_{x,MRO}$ unto the $(M,\mathbf{\hat{v}}_1, \mathbf{\hat{v}}_2)$ plane is perpendicular to $\mathbf{\hat{v}}_1$). Both situations would place the main lobe of an ideal dipole pointing towards the Sun. Lastly, the roll angle controls the rotation around the antenna axis. It thus affects the relative positions of SHARAD and the MRO bus (and other elements such as the high-gain antenna and the solar panels) with respect to the Sun. The gain of an ideal dipole is agnostic to roll, but here these large structures on the spacecraft are expected to play a role.

\begin{figure}[t]
    \centering
        \includegraphics[width=0.4\textwidth]{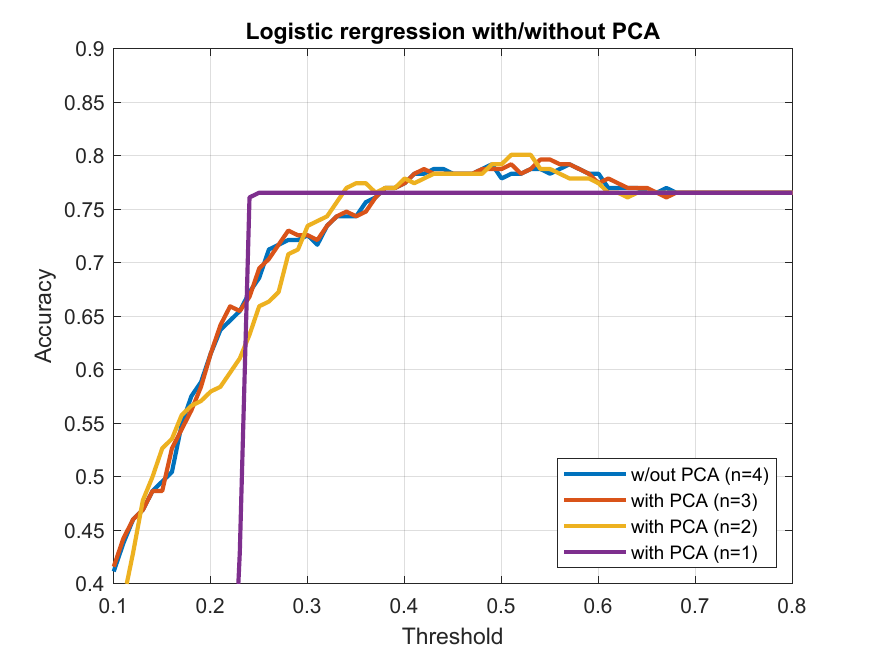}
    \caption{Multivariate logistic regression using the feature vectors generated by the principal component analysis.}
    \label{fig:pca}
\end{figure}

\begin{figure*}[t]
    \centering
        \subfloat[]{\includegraphics[width=0.4\textwidth]{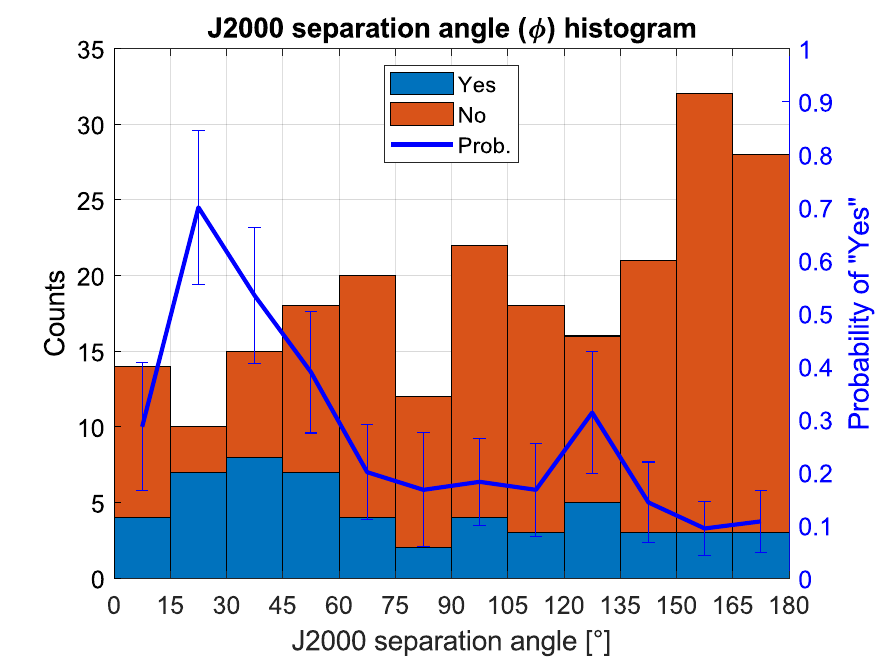}}
        \subfloat[]{\includegraphics[width=0.4\textwidth]{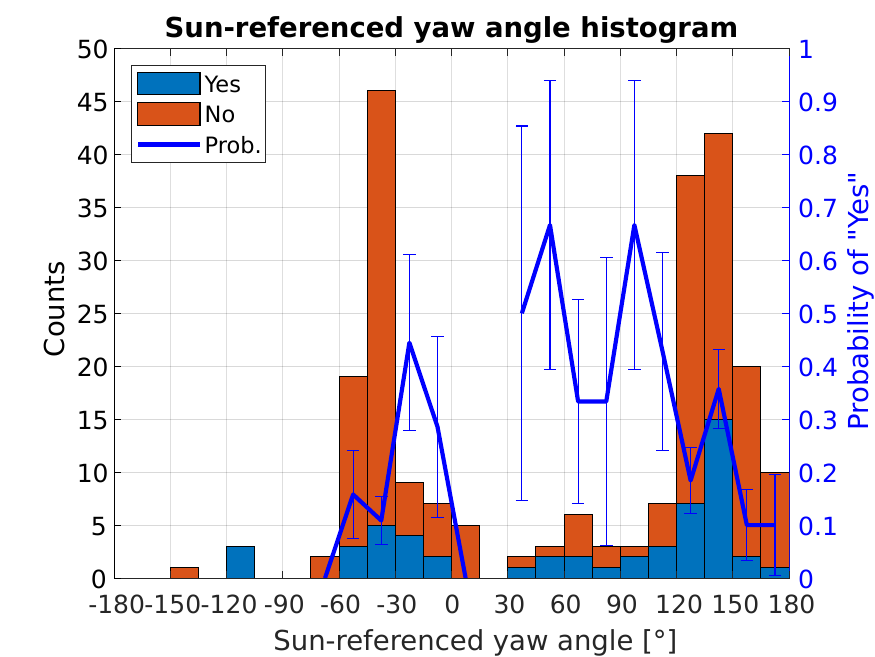}}\\
        \subfloat[]{\includegraphics[width=0.4\textwidth]{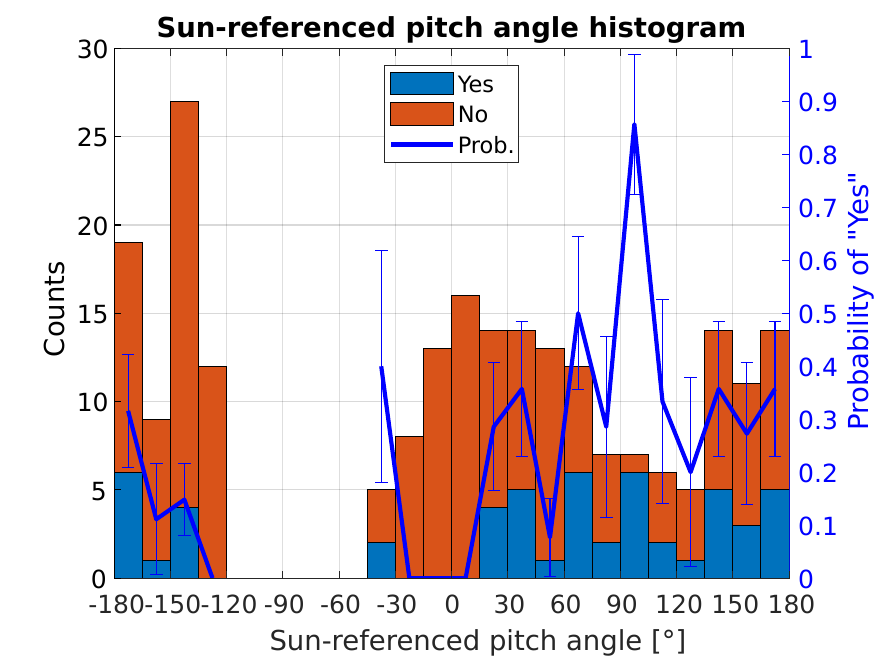}}
        \subfloat[]{\includegraphics[width=0.4\textwidth]{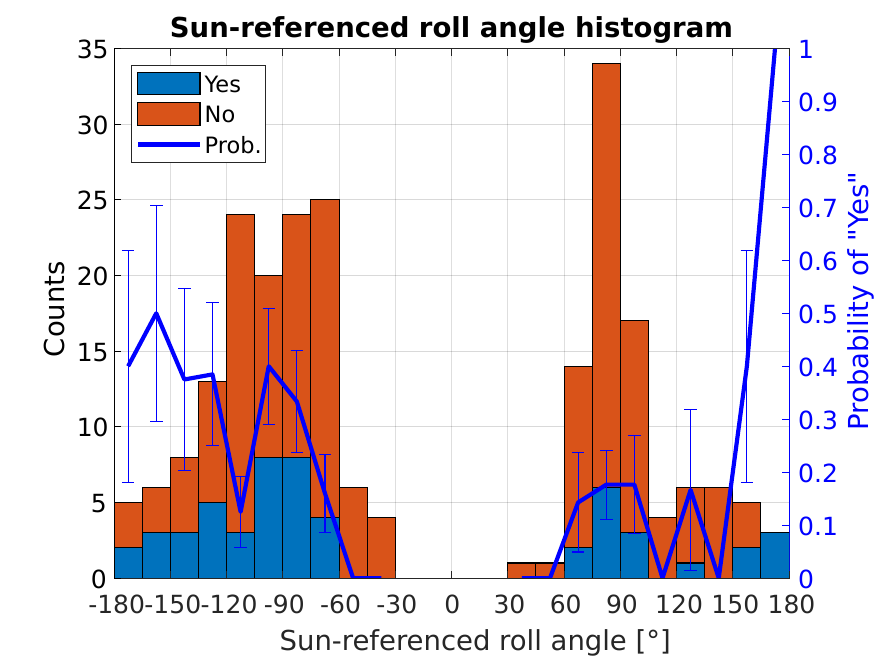}}
    \caption{Sensitivity of SHARAD to geometrical acquisition parameters. Histograms of the number of burst detections ("Yes") and burst no-detections ("No") along with the probability $p$ of detection (ratio of "Yes" to "Yes" + "No") for J2000 separation, yaw, pitch, and roll angles, respectively. The histograms bars are stacked, meaning the total number of events $n$ in a given angular bin is immediately accessible on the graph. The error bars associated with the probability curve correspond to the standard error of a binary outcome, that is, $\sigma = [p(1-p)/n]^{1/2}$.}
    \label{fig:angleshist}
\end{figure*}

We have made two types of statistical analysis based on the observer separation angle, MRO pitch, MRO yaw, and MRO roll: (i) a principal component analysis (PCA), and (ii) histograms and probability of detection for each angle independently. The results can be seen in Figure \ref{fig:pca} and \ref{fig:angleshist}a-d, respectively. 

Regarding the components of the PCA, which are orthogonalised linear combinations of the four angles, the first three contained 96.45\% of all variability in the original data, the first two 73.41\%, and the first component only 47.13\% of all variability. The correlations between any pair of angles is weak ($<0.2$) except for yaw and roll, which are strongly anti-correlated ($-0.8$), a result of the attitude control on MRO, itself on a polar orbit \citep{zurek2007overview}. We also performed a multivariate logistic regression using these PCA feature vectors to classify the bursts as “Yes” or “No”, based on a selectable threshold to convert the continuous output of the predictive model in the $[0,1]$ range to a “Yes”/“No” binary. This predictive model has a classification accuracy peaking at 0.792 for a threshold of 0.49 using all four feature vectors. A threshold near 0.5 is the most natural for an application such as this one; thus the fact that the logistic regression accuracy peaks at this value conforts the idea of a sensible model. Using only the first two components of the PCA, the predictive model retains similar accuracy, highlighting a degree of correlation between the four considered angles, as discussed. 

From the J2000 separation angle histogram, it is clear that detection is preferentially successful at low angles, with a probability of cross-detection reaching 0.7 when the angle between MRO and the source of observation is around $20^\circ$. Regarding orientation, the probability of detection peaks when pitch and yaw are both close to $90^\circ$. These angles put the radar axis perpendicular to the Sun (the former, vertically, and the latter, horizontally), which is the configuration for which maximal gain is expected. The absence of observations for some ranges of angles can be traced to MRO orbit specifics coupled with the no-occultation condition we imposed on the candidates. Probability of detection also peaks when the roll angle approaches $\pm 180^\circ$, that is, when $\mathbf{\hat{u}}_{z,MRO}$ is aligned but opposite to $\mathbf{\hat{v}}_1$ (see Figure \ref{fig:anglesschem}). For favourable yaw and pitch, such roll angles place SHARAD "in front" of the Sun, with the bus and the other large structures "behind" it. Intuitively, this is also a configuration that is expected to maximise the SHARAD gain, as it approaches that of traditional radar sounding \citep{campbell2021calibration}.

\section{Analysis and discussion} \label{sec:disc}
It is well understood that the morphology of the intensity-time profiles of a type III radio burst correspond to the growth of the instability \citep{Vosh15, Vosh15b, Krasnoselskikh19, Tkachenko21, Jebaraj23a} and the propagation of radio waves \citep{Fokker65, Arzner99}. The emission, which is a combination of all these processes, is systematically asymmetric in intensity-time profiles \citep{Suzuki85book}. Rapid variations in the peak intensity across the observing frequencies is not uncommon in decametre type III bursts \citep[][]{Kontar17}, and are to be expected when the electron beam evolves in an inhomogeneous medium. These variations may then be used to probe the plasma through which the beam propagates, assuming that the emission is close to the electron plasma frequency.

Figure \ref{fig:examples}a illustrates a typical type III burst observed on 22 December, 2011. STEREO A/WAVES also detected this burst, albeit at a lower resolution. Figure \ref{fig:examples}b displays a group of type III bursts observed on 01 January, 2012.
%The spectrogram broadening of type III bursts arises from the magnetized electrons following open magnetic field lines of the diverging solar magnetic field.
Due to limited resolution in the H-K wavelengths, different type III bursts within one are difficult to deconvolve. SHARAD, with its superior capabilities at lower frequencies, reveals nine bursts where STEREO A/WAVES only detects two. Figure \ref{fig:examples}c presents a type III burst with fine structures called \emph{striae} elements. These bursts can emerge due to electron beam evolution in inhomogeneous plasma, where stronger emissions occur in regions with smaller density inhomogeneities \citep{Jebaraj23a}. Pulse broadening is mainly due to velocity dispersion of the electron beam exciter \citep{reid2014review}, but the Langmuir wave growth can also be hindered in inhomogeneous plasma, resulting in less intense, longer-duration bursts \citep{Vosh15, Vosh15b}. In this example, SHARAD detects three distinct type III bursts, while STEREO A/WAVES observes a single burst without discernible fine structures, underscoring the significance of high time and frequency resolution.

While it is possible to distinguish fine structures and intensity variations using STEREO and Wind observations, their frequency resolution of $>$4\% would mean that only large scale inhomogeneities can be probed \citep[][]{Jebaraj23a}. On the other hand, SHARAD provides the resolution capable of resolving the fine-scale intensity variations which may then be used in tandem with modern missions such as the Parker Solar Probe during its close encounters. However, a complete characterisation of the spectral gain of SHARAD as well as its sources of EMI is likely to be needed for full exploitation of its spectral resolution. 

In this sense, the limitations of SHARAD for heliophysics are of two kinds. The first type of limitation are the instrument-specific limitations that we have just mentioned. Empirically, uneven spectral gain can be addressed by a de-trending of the time-average spectral power for each burst. Regarding EMI, notch filtering has been applied successfully to SHARAD products to enhance the quality of range-compressed observations \citep{campbell2021calibration}. While these methods are promising for heliophysics-oriented analysis of SHARAD data, one must ensure they do not compromise the natural signals of interest. For this reason, we devolve the study of fine structure within the very high-resolution SHARAD dynamic spectra of type-III bursts to a future study. The second type of limitation arises from using SHARAD opportunistically: the limited number of bursts that we observe in our dataset can be traced to the low duty cycle of SHARAD, the preference for night-time observations for radar sounding of Mars, and the nadir-pointing geometry favoured for radar sounding of Mars.

\begin{figure}[t]
    \centering
    \includegraphics[width=0.4\textwidth]{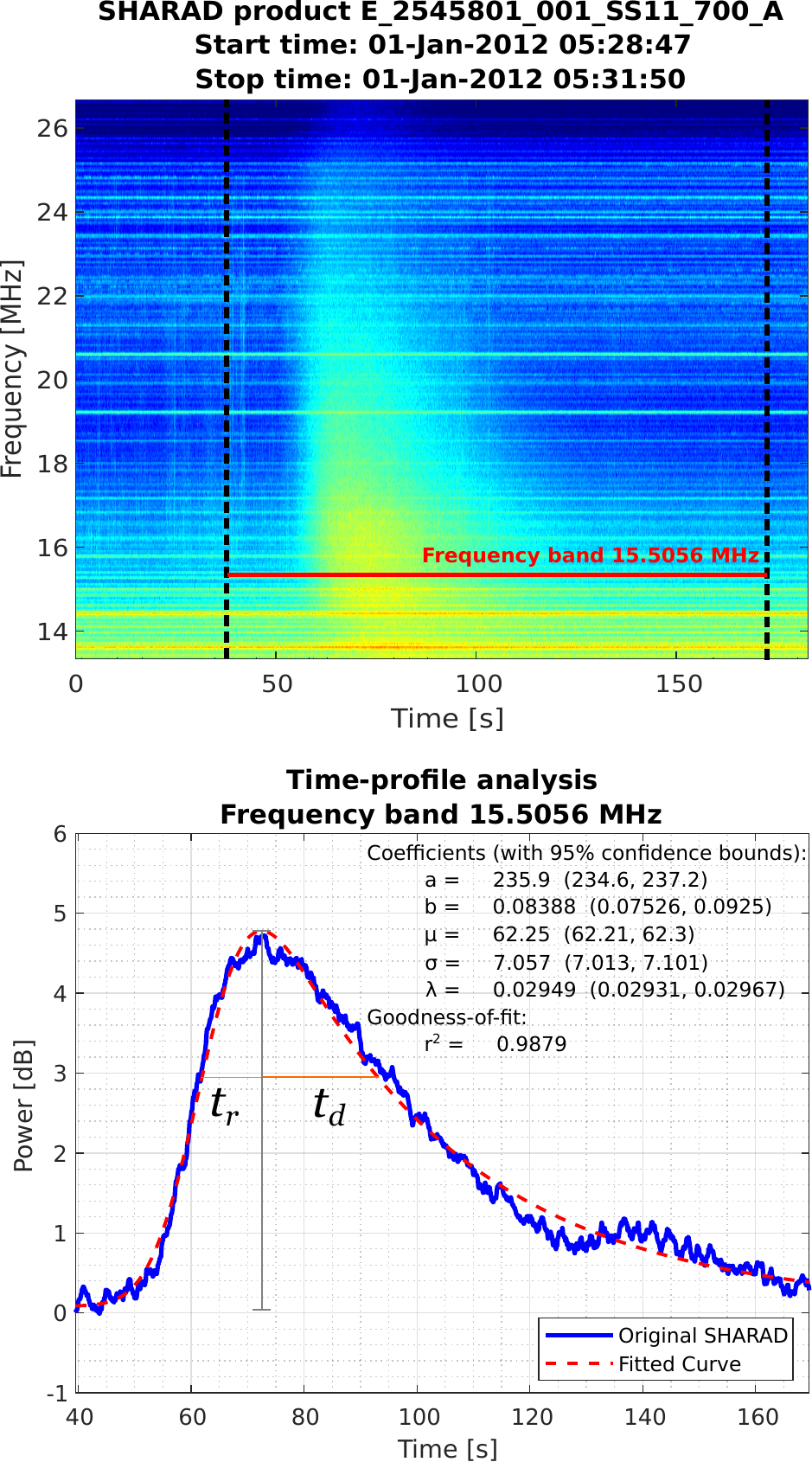} 
    \caption{Worked-out example of a time-profile extraction and fit on SHARAD product 2545801 using the EMI-free frequency band of 15.5056 MHz. The black vertical dashed lines on the dynamic spectrum represent the portion of the product that was fed into the fitting subroutine. The coefficients displayed next to the fitted curve correspond directly to those of equation \ref{exgaussian}. The rise and decay times $t_r$ and $t_d$ were derived as the first and second half-width at half-maximum, itself computed as the difference between the maximum of the curve and the floor parameter $b$.}
    \label{fig:fitexample}
\end{figure}

\subsection{Time-profile characteristics at 1.5 AU}\label{sec:risefall}

The very high temporal resolution of SHARAD has the consequence that the fast Gaussian-like rise and the slower exponential decay are effortless to identify, as evidenced by the examples shown in Figure \ref{fig:examples}. For this reason we propose to study the statistics of the frequency-dependence of the rise time $t_r$ and decay time $t_d$, computed by fitting an exponentially-modified Gaussian function \citep[][]{Dulk84, reid2018solar, jebaraj2023c} and by computing the half-widths at half-maximum (HWHM) on either side of the peak intensity. \cite{reid2018solar} conducted a comparable study with 31 type-III bursts recorded by LOFAR \citep{VanHaarlem13} in the 30 to 80 MHz range. The characteristic times of type-III bursts in the frequency range of 15 to 25 MHz, which provides a natural continuation to \citep{reid2018solar}, has not been explored beyond 1 AU at the resolutions we have access to here, and constitute a novel result. 

The time-profile analysis was run on all the "canonical" type-III bursts detected by SHARAD, that is, all those listed in Table \ref{tab:yesbursts} with an asterisk in the "Notes" column. This excludes those with peculiar morphologies (\emph{e.g.} 2712401), those with extensive fine-structure (\emph{e.g.} 3617602), and those which are not type-III's (\emph{e.g.} 3544101). A similar sieving was done in \cite{reid2018solar}. When a given SHARAD data product contained several bursts, all those that are exploitable within this product were included. This left 26 individual bursts to analyse. Of these bursts, all frequency lines containing EMI were ignored. For a given product, the EMI lines were identified through a global time-average of the dynamic spectrum, the application of a median filter with kernel size 101, and the detection of all frequency lines that were above this trendline. We also exclude the frequency lines at the edges of the instrument's capabilities, and ran our analysis in the 15 -- 25 MHz range, the bandwidth the instrument was optimised for. After exclusion of EMI and edge frequencies, about 800 frequency lines per burst (out of 1800) are still available for time-profile analysis. In total, we thus analysed about 25\,000 profiles.

The function resulting from the convolution of an exciter function and an exponential decay is able to accurately represent the intensity time-profiles of type-III bursts \citep{aubier1972,barrow1975solar}. With the choice of a Gaussian function as the exciter \citep{degroot1966}, this function takes the name of exponentially-modified Gaussian (EMG) function. It is more general  than both the Gaussian used in \cite{reid2018solar} and the exponential decay functions historically used to study type III bursts \citep{evans1973characteristics,dulk2000type}, and yields particularly accurate fits in our study of lower-frequency bursts due to its asymmetry. The characteristic times extracted from this function will therefore be the most faithful. The five-parameter function we use to fit the time-profiles of this paper is given by:
\begin{align} \label{exgaussian}
        g(a,b,\mu,\sigma,\lambda; x) =& a \int_0^x  \frac{1}{\sigma\sqrt{2\pi}}  e^{-\frac{1}{2}\left(\frac{x-\bar{x}-\mu}{\sigma}\right)^2} e^{-\lambda \bar{x}} d\bar{x} + b \\
     \begin{split}
       =& a \frac \lambda 2 \text{exp}\left\{ \frac \lambda 2 \left( 2\mu + \lambda\sigma^2 - 2x\right) \right\}\\
        &\left( 1- \text{erf}\left\{ \frac{\mu+\lambda\sigma^2 - x}{\sqrt{2}\sigma}\right\}\right) + b,      
    \end{split}
\end{align}
where $a$ controls the overall scaling of the burst, and $b$, its floor; $\mu$ is the mean of the Gaussian sector of the function, and $\sigma^2$, its variance; lastly, $\lambda$ controls the decay rate of the exponential sector. The error function is defined as $\text{erf}(z) \equiv 2(\pi)^{-1/2} \int_0^z e^{-t^2}\text{d}t$. The rise time $t_r$ and decay time $t_d$ are then computed as the two half-widths at half-maximums of the fitted curve. In Figure \ref{fig:fitexample} we show an example of a SHARAD profile along with its fitted function, and a graphical representation of $t_r$ and $t_d$. 

Along with the quantities of interest, we also recorded the $r^2$ coefficient for each fit, in order to exclude fits of bad quality. Bad fits can occur when the signal-to-noise ratio of the burst is very low, when several closely-spaced bursts forces us to constrain the cropping around each of them, or when the profile is contaminated by reflections of the active SHARAD chirp. Due to the abundance of data points, we were able to apply very strict fit quality criteria, considering only the profiles for which $r^2>0.95$, and still be left with about 7000 data points in the final analysis.

\begin{figure*}[h!t]
    \centering
        \includegraphics[width=0.85\textwidth]{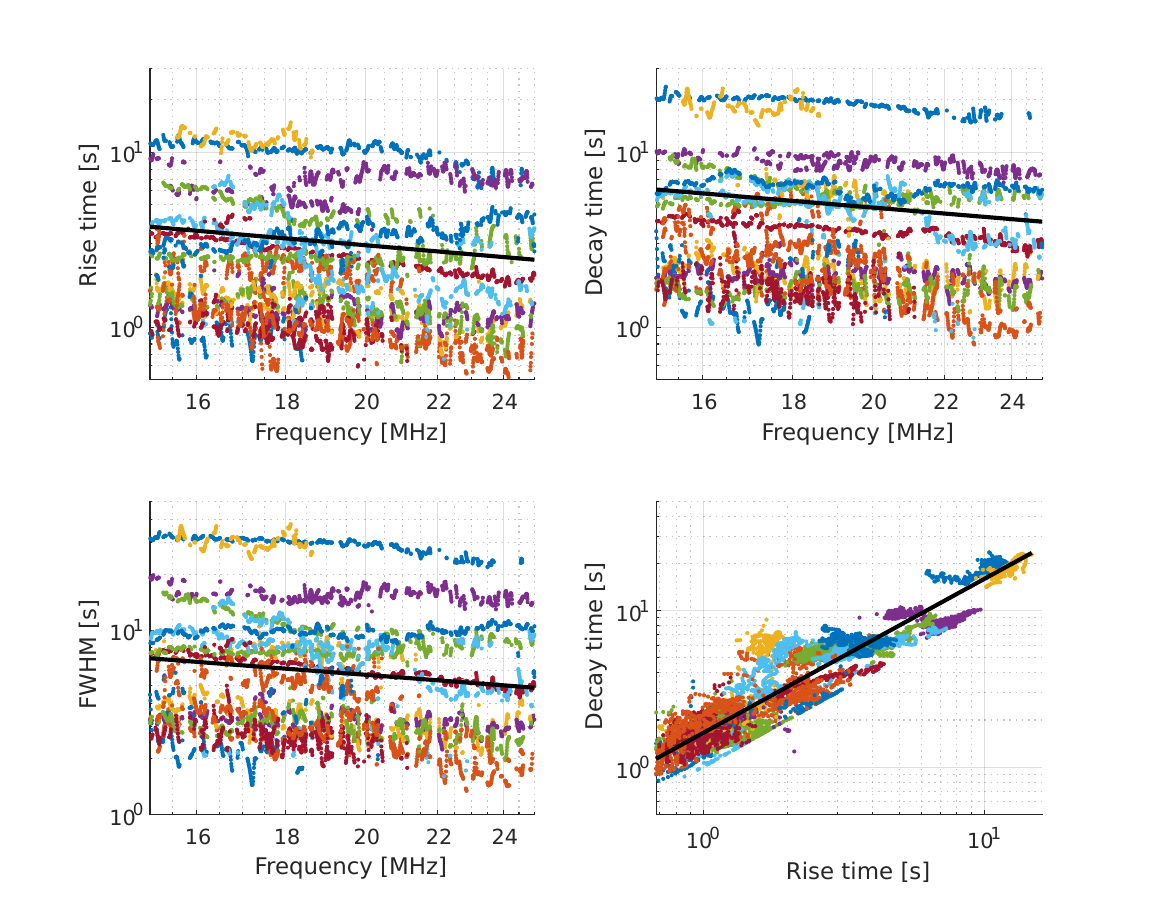}
    \caption{Analysis of rise time and decay time of the suitable type-III bursts recorded by SHARAD after fitting an exponentially-modified Gaussian function onto every frequency line that is unhampered by EMI. The color-coding of the dots represents profiles from an individual burst, with some colours being reused. The black line is the best power-law fit, the equations for which is given in formulae \ref{tr} through \ref{tftr}.}
    \label{fig:risefall}
\end{figure*}

The results for $t_r(f)$, $t_d(f)$, and the derived quantities $\text{FWHM}(f) = (t_d+t_r)(f)$ and $t_d(t_r)$ can be seen in Figure \ref{fig:risefall}, and the best fit power laws for each plot are given in equations \eqref{tr} through \eqref{tftr}, respectively. These laws were obtained by directly fitting a power function using MATLAB's unweighted non-linear fit method, rather than by linearly fitting the data in logarithm space. This was done so as (i) to minimise the non-linear transformation applied on to the data, and (ii) to conform with the procedure of \cite{reid2018solar}.
\begin{eqnarray}
t_r &=& \left(38\pm16\right) \left( \frac{f \text{[MHz]}}{1 \text{[MHz]}} \right)^{-0.85 \pm 0.14} \label{tr}\\
t_d &=& \left(58\pm24\right)  \left( \frac{f \text{[MHz]}}{1 \text{[MHz]}} \right)^{-0.83 \pm 0.14} \label{td}\\
\text{FWHM} &=&  \left(97\pm40\right)   \left( \frac{f \text{[MHz]}}{1 \text{[MHz]}} \right)^{-0.84 \pm 0.14}\label{d}\\ 
t_d &=& \left(1.654\pm0.025\right)  \left( \frac{t_r \text{[s]}}{1 \text{[s]}} \right)^{0.985 \pm 0.007}\label{tftr}
\end{eqnarray}

\subsubsection{Rise and decay time}

Both the SHARAD rise time exponent $-0.85 \pm 0.14$ (eq. \ref{tr}) and the SHARAD decay time exponent $-0.83 \pm 0.14$  (eq. \ref{td}) are consistent with \cite{reid2018solar}, who found $-0.77 \pm 0.14$ and $-0.89 \pm 0.15$, respectively, using LOFAR. Interestingly, \citep{reid2018solar} finds the decay time to decrease more sharply with frequency, whereas we find similar behaviour. The rise time is closely linked to the exciter function, \emph{i.e.} the growth of Langmuir waves and inherits its characteristics \citep{Krasnoselskikh19}, and it is thus expected to behave similarly at different points of the solar system for a given frequency. The decay time, on the other hand, is not completely immune to propagation effects, and can be thought as a convolution between the characteristics of the source and the electromagnetic wave propagation effects \citep{reid2018solar} ; it is therefore not excluded that the subtle discrepancy we observe for the decay time at 1.5 AU is partly due to such effects, although it seems unlikely given the frequency range of SHARAD. On the other hand, it is widely understood that these effects are minimal for higher harmonics \citep{Melrose1980}. Recent studies by \cite{jebaraj2023c} have found that type III bursts predominantly occur as fundamental-harmonic pairs. We did not differentiate between fundamental and harmonic components in this study, so we cannot exclude some kind of selection bias towards the former, either. At any rate, these broadly-consistent results can be interpreted as validation of the appropriateness of SHARAD, but a thorough investigation of the fundamental vs. harmonic content of our SHARAD-detected burst dataset should yield interesting insights. We defer this to a future study.

\subsubsection{Burst duration}
\begin{figure}[t]
    \centering
    \includegraphics[width=0.4\textwidth]{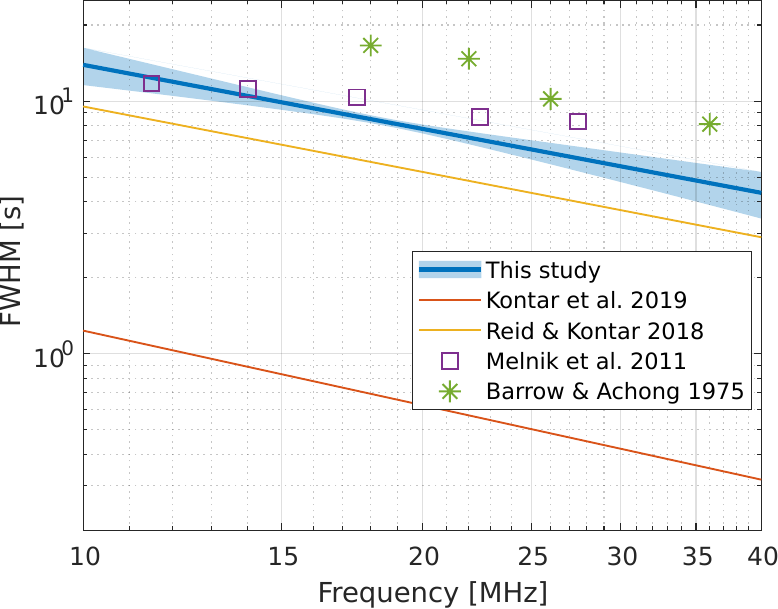} 
    \caption{\label{fig:fwhm} Comparison of the type III burst durations (FWHM) derived from SHARAD data (eq. \ref{d} and Figure \ref{fig:risefall}) with that of \citep{kontar2019anisotropic, reid2018solar, melnik2011observations,barrow1975solar}.The shaded area around the law obtained in this study corresponds to the 95\% confidence interval for the parameters of the fitted function (functional prediction interval).}
\end{figure}

The burst duration is commonly computed as the full width at half-maximum (FWHW) of the time profile in the literature \citep{barrow1975solar, reid2018solar, kontar2019anisotropic}. In our case, this quantity is readily available as the sum of the fall and decay times, FWHM$= t_r + t_d$. Our duration values are somewhat higher than that suggested by \cite{kontar2019anisotropic}, whose study examines a much wider range of frequencies (100 kHz -- 100 MHz), but comparable than that of \cite{melnik2011observations}, who studied particularly powerful bursts in a similar frequency range as our. Duration of SHARAD bursts is, however, slightly shorter than those detected by \cite{barrow1975solar} with the University of West Indies Radio Observatory. Our analysis yields a frequency-dependence with an exponent of $-0.84 \pm 0.14$ (eq. \ref{d}), which is compatible with that of both \cite{kontar2019anisotropic} and \cite{reid2018solar}, who found $-0.98\pm0.05$ and $-0.86\pm0.11$. A graphical summary of these comparisons can be found in Figure \ref{fig:fwhm}.

\subsection{Burst asymmetry}
Computing the decay time as a function of the rise time (eq. \ref{tftr}), we find that the former is almost always longer than the latter, as expected, outliers notwithstanding. The exponent of its frequency-dependence close to one also matches with the literature \citep{reid2018solar}.

 \section{Conclusions and outlook} \label{sec:concl}
In this manuscript, we demonstrated that planetary radar sounders such as SHARAD can be used as high-resolution solar radio-observatories, we presented a quantitative analysis of the sensitivity of SHARAD to solar burst events based on the statistics of correlated observations from a geometric standpoint, and extracted the characteristic times of type-III bursts from this dataset.

Our findings show that SHARAD observations can bring the large advancements in the understanding of the generation and propagation of the radio bursts at the distances even as far as the Mars orbit. In addition to studying characteristic times at 1.5 AU, the capabilities of SHARAD could be leveraged to study fine-structure, provide additional points of observation in burst triangulation, and its bandwidth completes the spectrum between typical dedicated solar missions and earth radiotelescopes. From the perspective of radar science, correlated observations with STEREO can also help the absolute calibration of SHARAD since STEREO/WAVES data is absolutely-calibrated: since bursts can be compared one-to-one between the observer(s) and SHARAD, we could predict the power that SHARAD should have detected for a given burst, taking into account an estimation of the directional gain of SHARAD (for instance, through EM simulations) and the effects of longitudinal variations (for instance, by selecting situations where the observer and MRO are aligned, or by a method such as that of \cite{Musset21}). Conversely, it is also possible to start from current indirect calibrations of SHARAD to estimate the absolute power of detected bursts.

It must be noted that the catalogue of type III bursts we present is limited to correlated observations only. There are likely more bursts within the SHARAD archive that could be found through direct inspection. For example, a survey of the MARSIS dataset is planned amongst future work in order to produce a complete catalogue of MARSIS solar radio burst observations (Sánchez-Cano et al., in prep.), which will be compared with SHARAD to get a fuller frequency spectrum of the co-observed bursts. The main limitation of using radar sounder data of the Sun is the fact they typically operate no more than a few minutes or tens of minutes a day, and cannot function as a round-the-clock survey instrument.

The results presented here can also be used for future planetary missions. In this decade, two missions carrying radar sounder instruments are being launched towards the Jupiter system: the Radar for Icy Moons Exploration (RIME) onboard the ESA Jupiter Icy Moons Explorer (JUICE) probe, and the Radar for Europa Assessment and Sounding Ocean to Near-Surface (REASON) onboard the NASA Europa Clipper mission \citep{reason,rime} ; as well as one towards Venus : the Subsurface Radar Sounder (SRS) onboard the ESA Envision mission \citep{srs}. Maturing the scientific exploitation of solar bursts with SHARAD opens the way to utilise these new instruments as solar observatories as well.

\begin{acknowledgements}
Christopher Gerekos dedicates this manuscript to the memory of his beloved mother Patricia Brabants (1965--2023). This work was supported by the G. Unger Vetlesen Foundation and by JPL’s Innovative Spontaneous Concepts in Research and Technology Development program. Part of this research was carried out at the Jet Propulsion Laboratory, California Institute of Technology, under a contract with the National Aeronautics and Space Administration. I.C.J. is grateful for support by the Academy of Finland (SHOCKSEE, grant No.\ 346902). B.S.-C. acknowledges support through STFC Ernest Rutherford Fellowship ST/V004115/1. M.L. acknowledges support through STFC grant ST/W00089X/1. J.M. acknowledges funding by the  BRAIN-be project SWiM (Solar Wind Modeling with EUHFORIA for the new heliospheric missions). The authors wish to express their gratitude to Bruce Campbell, Dirk Plettemeier, Marco Mastrogiuseppe, Vratislav Krupar, Marc Pulupa, Vladimir Krasnoselskikh, and Milan Maksimovic for the excellent discussions.
\end{acknowledgements}

%\bibliographystyle{aa}
%\bibliography{mybibfile} 

\appendix

\section{SHARAD duty cycle analysis} \label{sec:dutycyle}

To provide additional context on the discussion of the general properties of the dataset and to highlight the targetted nature of the instrument (as opposed to a survey instrument such as WAVES), we present a brief analysis of the activity duty cycle of SHARAD over the years. Starting from the SHARAD EDR label files, we extracted the radargram start time and stop time for each radargram in the entire SHARAD dataset in the period of study (6 December 2006 to 31 December 2021) in order to compute the duration of that radargram. Taking the cumulative sum of all these durations for each year, we obtain the durations shown in Table \ref{tab:dutycycle}. On an average year, SHARAD will have been on for about a million seconds, corresponding to a duty cycle of about 3\%. Aside from the solar cycle, part of the variability of the number of candidates counted in Figure \ref{fig:generals}a can be traced to the fraction of the time SHARAD was operating (for instance, the dip in candidates in 2013).

\begin{table}[h]
    \centering
    \begin{tabular}{c|c c c}
        & Activity [$10^6$ s] & Duty cycle [\%] \\
        \hline \hline
        2006* & 0.05 & 0.15 \\
        2007 & 1.01 & 3.48 \\
        2008 & 1.14 & 3.60 \\
        2009 & 0.48 & 1.51 \\
        2010 & 1.06 & 3.36 \\
        2011 & 0.84 & 2.67 \\
        2012 & 1.35 & 4.28 \\
        2013 & 0.68 & 2.17 \\
        2014 & 1.50 & 4.74 \\
        2015 & 0.42 & 1.32 \\
        2016 & 1.08 & 3.43 \\
        2017 & 0.60 & 1.91 \\
        2018 & 1.08 & 3.43 \\
        2019 & 0.77 & 2.44 \\
        2020 & 1.74 & 5.51 \\
        2021 & 1.07 & 3.39 \\
    \end{tabular}
    \caption{SHARAD operating time throughout the surveyed years, expressed in number of seconds per year (activity) and in percentage over a given year (duty cycle). The asterisk (*) denotes a partial year.}
    \label{tab:dutycycle}
\end{table}

\clearpage

\vspace{20cm}

\newpage

\section{Quicklook of SHARAD SRB products}\label{sec:quicklook}
\vspace{-0.4cm}
\begin{figure}[h!]
\centering
\includegraphics[width=0.92\textwidth]{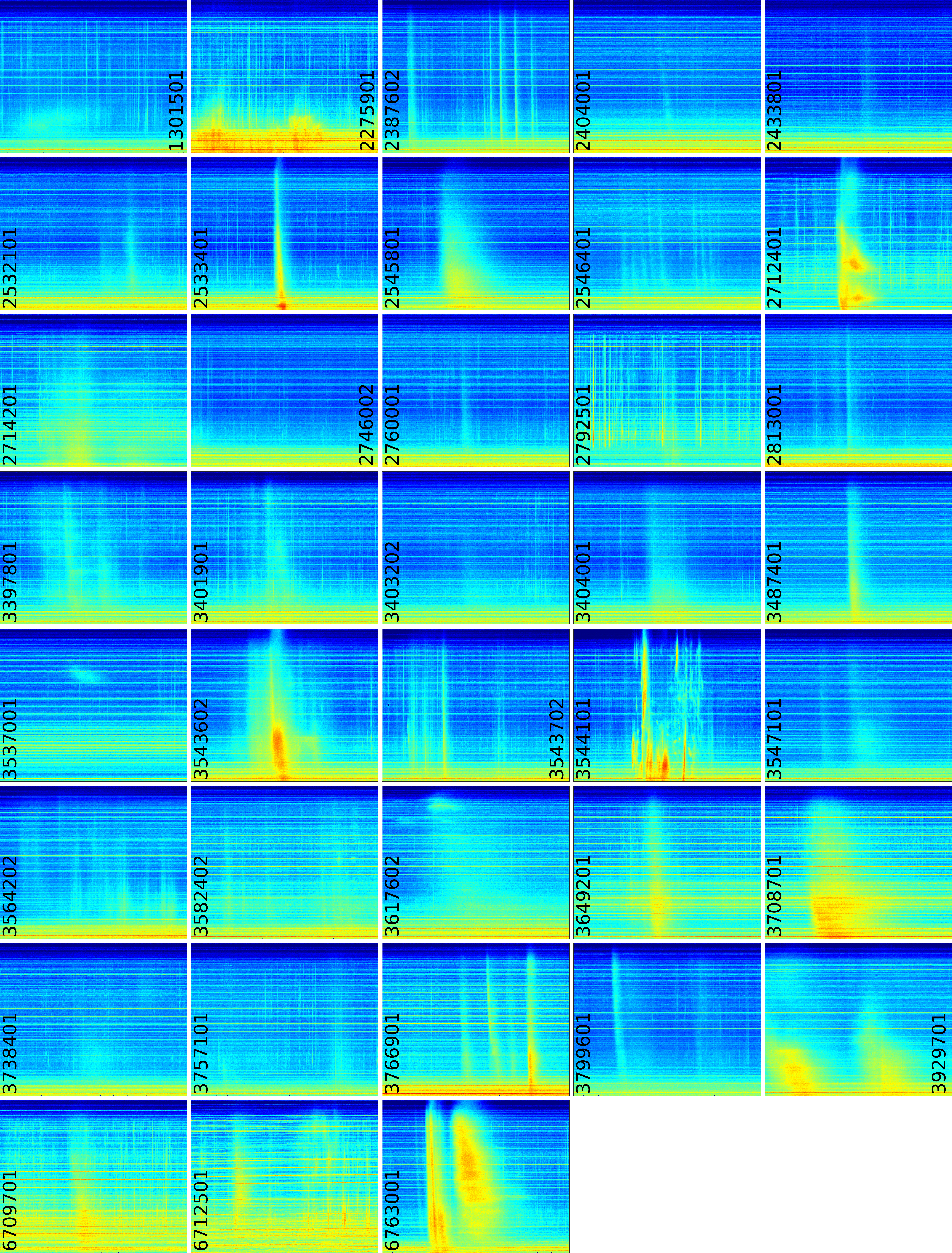}
\caption{Dynamic spectra of all SHARAD-detected solar burst products. The y-axis goes from 13.3 to 26.7 MHz for all images. The x-axis bounds is specified in Table \ref{tab:yesbursts}. The dynamic range of the image can vary from acquisition to acquisition.}
\end{figure}

\end{document}